\documentclass[fleqn,12pt]{wlscirep}
\usepackage{graphicx}
\usepackage{bm}

\title{Transition between strong and weak topological insulator in ZrTe$_5$ and HfTe$_5$}

\author[1]{Zongjian Fan}
\author[2]{Qi-Feng Liang}
\author[3]{Y. B. Chen}
\author[1]{Shu-Hua Yao}
\author[1,4,*]{Jian Zhou}
\affil[1]{National Laboratory of Solid State Microstructures and Department of Materials Science
                and Engineering, Nanjing University, Nanjing 210093, China }
\affil[2]{Department of Physics, Shaoxing University, Shaoxing 312000, China}
\affil[3]{National Laboratory of Solid State Microstructures and Department of Physics,
                Nanjing University, Nanjing 210093, China}
\affil[4]{Collaborative Innovation Center of Advanced Microstructures, Nanjing University, Nanjing , 210093, China.}
\affil[*]{Corresponding author: zhoujian@nju.edu.cn}

\begin{abstract}
ZrTe$_5$ and HfTe$_5$ have attracted increasingly attention recently since the theoretical prediction of being topological
insulators (TIs). However, subsequent works show many contradictions about their topological nature.
Three possible phases, i.e. strong TI, weak TI, and Dirac semi-metal, have been observed in different experiments until now.
Essentially whether ZrTe$_5$ or HfTe$_5$  has a band gap or not is still a question.
Here, we present detailed first-principles calculations on the electronic and topological properties of ZrTe$_5$ and HfTe$_5$
on variant volumes and clearly demonstrate the topological phase transition from a strong TI, going through an intermediate
Dirac semi-metal state, then to a weak TI when the crystal expands.
Our work might give a unified explain about the divergent experimental results and
propose the crucial clue to further experiments to elucidate the topological nature of these materials.
\end{abstract}

\begin{document}

\flushbottom
\maketitle
\thispagestyle{empty}

\section*{Introduction}

Topological insulator (TI) is a new class of material which is an insulator in its
bulk, while having time reversal symmetry protected conducting states on the
edge or surface.  ~\cite{rev1,rev2,rev3}  A large number of realistic materials have
been theoretically proposed and experimentally confirmed, such as Bi$_2$Se$_3$ and Bi$_2$Te$_3$.~\cite{bise1,bise2}
However, the layered transition-metal pentatelluride ZrTe$_5$ and HfTe$_5$ is a particular example.
ZrTe$_5$ and HfTe$_5$ were studied more than 30 years ago due to the large thermoelectric power ~\cite{thermo}
and mysterious resistivity anomaly.~\cite{resistivity1,resistivity2}
Recently, Weng \textit{et al.}  predicted that mono-layer ZrTe$_5$ and HfTe$_5$
 are good quantum spin Hall insulators with relatively large bulk band gap (about 0.1 eV)
 by first principles calculations.~\cite{wenghm1}
The three-dimensional (3D) bulk phase of ZrTe$_5$ and HfTe$_5$  are also predicted to be TIs,
 which are located at the vicinity of a transition between strong and weak TI,
 but without detailed description.~\cite{wenghm1}

Nevertheless,  subsequent experiments show many contradictions about the topological nature of ZrTe$_5$ or HfTe$_5$.
Several experimental works suggested that  ZrTe$_5$ is a Dirac semi-metal without a finite band gap
 by different characterization methods, such as Shubnikov-de Haas oscillations, angle-resolved photoemission
 spectroscopy (ARPES), and infrared reflectivity measurements.~\cite{chen1,chen2,chiral,yliu,xiangyuan,glzheng}
Of course there are also other experimental works holding opposite point of view.
For example, in two recent scanning tunneling microscopy (STM) experiments,
they unambiguously observed a large bulk band gap about 80 or 100 meV  in ZrTe$_5$, ~\cite{lisc1,wenghm2}
implying that there is no surface state on the top surface and therefore ZrTe$_5$ should be a weak TI.
Another APRES work also favored a weak TI for ZrTe$_5$.~\cite{moreschini1}
However, there are two other ARPES works which believed that ZrTe$_5$ is a strong TI.  ~\cite{manzoni1,manzoni2}
For instance, by using the comprehensive ARPES, STM, and first principles calculations, Manzoni \textit{et al.}
 found a metallic density of state (DOS) at Fermi energy, which arises from the two-dimensional surface state and thus indicates
  ZrTe$_5$ is a strong TI. ~\cite{manzoni1}

The divergence of these experiments make ZrTe$_5$ (HfTe$_5$) being a very puzzling but interesting material,
which needs more further experimental and theoretical studies. Therefore, in order to figure out the physical
mechanism behind those contradictory experimental results, we revisited the band structures of ZrTe$_5$ and HfTe$_5$,
and carefully studied their relationship with the volume expansion. We find a clear topological transition
between a strong and weak TI in ZrTe$_5$ and HfTe$_5$, accompanied by an intermediate Dirac semi-metal state between them.
This work could shed more light on a unified explain about the different experimental results, and
propose the crucial clue to further experiments to elucidate the topological nature of  ZrTe$_5$ and HfTe$_5$.

\section*{Results}

As shown in Fig. 1(a), ZrTe$_5$ and HfTe$_5$ share the same base-centered orthorhombic crystal structure
with Cmcm (No. 63) space group symmetry. Trigonal prismatic ZrTe$_3$ chains oriented along the $a$ axis
make up the ZrTe$_5$ natural cleavage plane. Each chain consists of one Zr atom
and two different kinds of Te atoms. ZrTe$_3$ chains are connected by zigzag Te chains along the $c$ axis,
building a two dimensional structure of ZrTe$_5$ in the $a$-$c$ plane.
One crystal unit cell contains two ZrTe$_5$ planes piled along the $b$ axis, the stacking orientation of ZrTe$_5$.
The Brillouin zone and its high symmetry k-points of ZrTe$_5$ (HfTe$_5$) are shown in Fig. 1(b), in which
$a^*$, $b^*$, and $c^*$ are the reciprocal lattice vectors.

\begin{figure}[ht]
\centering
\includegraphics[width=0.7\linewidth]{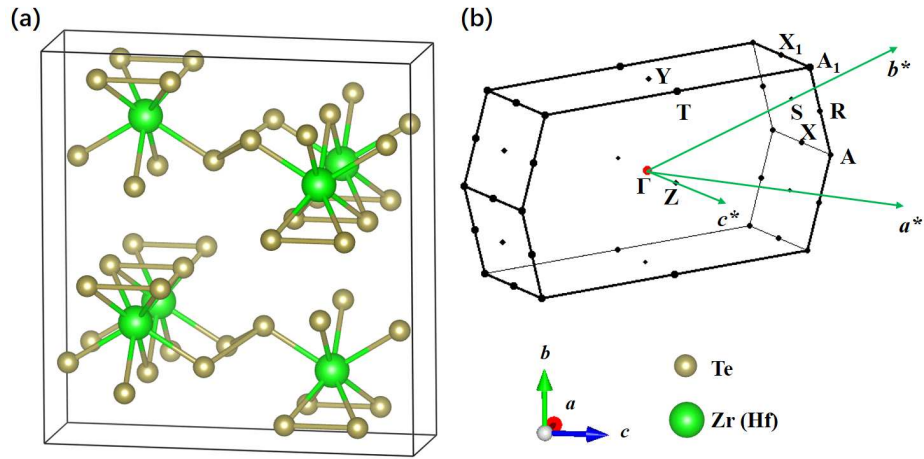}
\caption{\label{fig:str} (Color online) ({\bf a}) Layered crystal stucture of ZrTe$_5$ (HfTe$_5$)  in the orthorhombic
conventional unit cell. Big green and small brown balls represent Zr (Hf) and Te atoms respectively.
The layers stack along the $b$ direction.
({\bf b}) Brillouin zone  and the high symmetry points of  the primitive unit cell of ZrTe$_5$ (HfTe$_5$).  }
\end{figure}

\begin{table}[ht]
\caption{ Calculated and experimental lattice constants  and volumes of
ZrTe$_5$ and HfTe$_5$ in conventional unit cell. Experimental data is
 from reference 21.  }
\begin{center}
\begin{tabular}{c|c|c|c|c|c} \hline\hline
      Material       &  Method  &  $a$  (\AA)  & $b$  (\AA)  & $c$  (\AA)  & $V$   (\AA$^3$)           \\ \hline
                     &  PBE                 & 4.0490 & 15.772 & 13.845 & 884.17  \\  \cline{2-6}
                     &  optB86b-vdw & 4.0064 & 14.590 & 13.732 & 802.69  \\  \cline{2-6}
    ZrTe$_5$  & exp.  (293 K)  & 3.9875 & 14.530 & 13.724 & 795.15       \\ \cline{2-6}
                     &  exp. (10 K)   & 3.9797 &  14.470 & 13.676 & 787.55    \\   \hline
                     &    PBE                  & 4.0245 & 15.694 & 13.843 & 874.37 \\  \cline{2-6}
                     &    optB86b-vdw   & 3.9799 & 14.564& 13.743 &  796.58      \\ \cline{2-6}
    HfTe$_5$    &   exp. (293 K) & 3.9713 & 14.499 & 13.729 & 790.51   \\ \cline{2-6}
                     &   exp. (10 K)  & 3.9640 & 14.443 & 13.684 & 783.44   \\  \hline \hline
\end{tabular}
\end{center}
\end{table}

\begin{figure}[ht]
\centering
\includegraphics[width=0.5\linewidth]{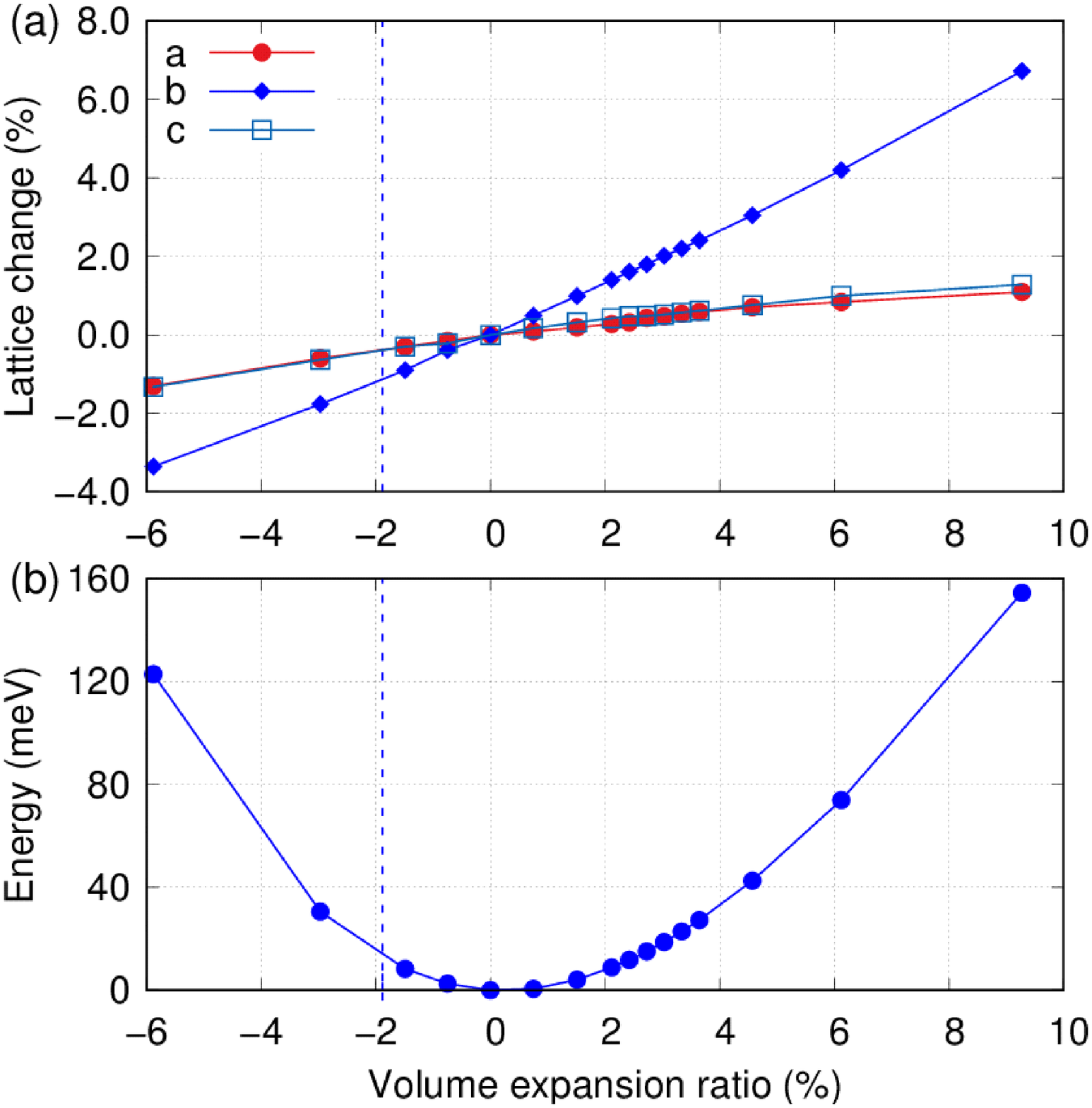}
\caption{\label{fig:str} (Color online) ({\bf a})   Lattice constants $a$, $b$, and $c$ under different unit cell volumes.
 ({\bf b})  Relative total energy of a primitive unit cell of ZrTe$_5$ under different volumes.
 Blue vertical dotted lines represent the experimental volume at 10 K. ~\cite{explattice}
 }
\end{figure}

Due to the weak the van der Waals (vdw) interaction in the layer ZrTe$_5$ (HfTe$_5$),~\cite{wenghm1}  the vdw corrected
correlation functional is necessary in order to obtain the good theoretical lattice constants. In Table I,
we present the optimized lattice constants of ZrTe$_5$ and HfTe$_5$ based on the optB86b-vdw functional,
as well as the experimental ones. ~\cite{explattice} We find that the theoretical and
experimental lattice constants are well consistent with each other and the maximum difference between
them is less than 1\%. For comparison, we also optimized the structures of  ZrTe$_5$ and HfTe$_5$ by using the standard
Perdew-Burke-Ernzerhof (PBE) exchange correlation. ~\cite{pbe}  It is obvious that there is a large
error (about 9 \%) in the lattice $b$ (in the stacking direction), which indicates that standard PBE failed
to describe the structures of ZrTe$_5$ andHfTe$_5$, and the vdw correction is necessary.

\begin{figure}[ht]
\centering
\includegraphics[width=0.9\linewidth]{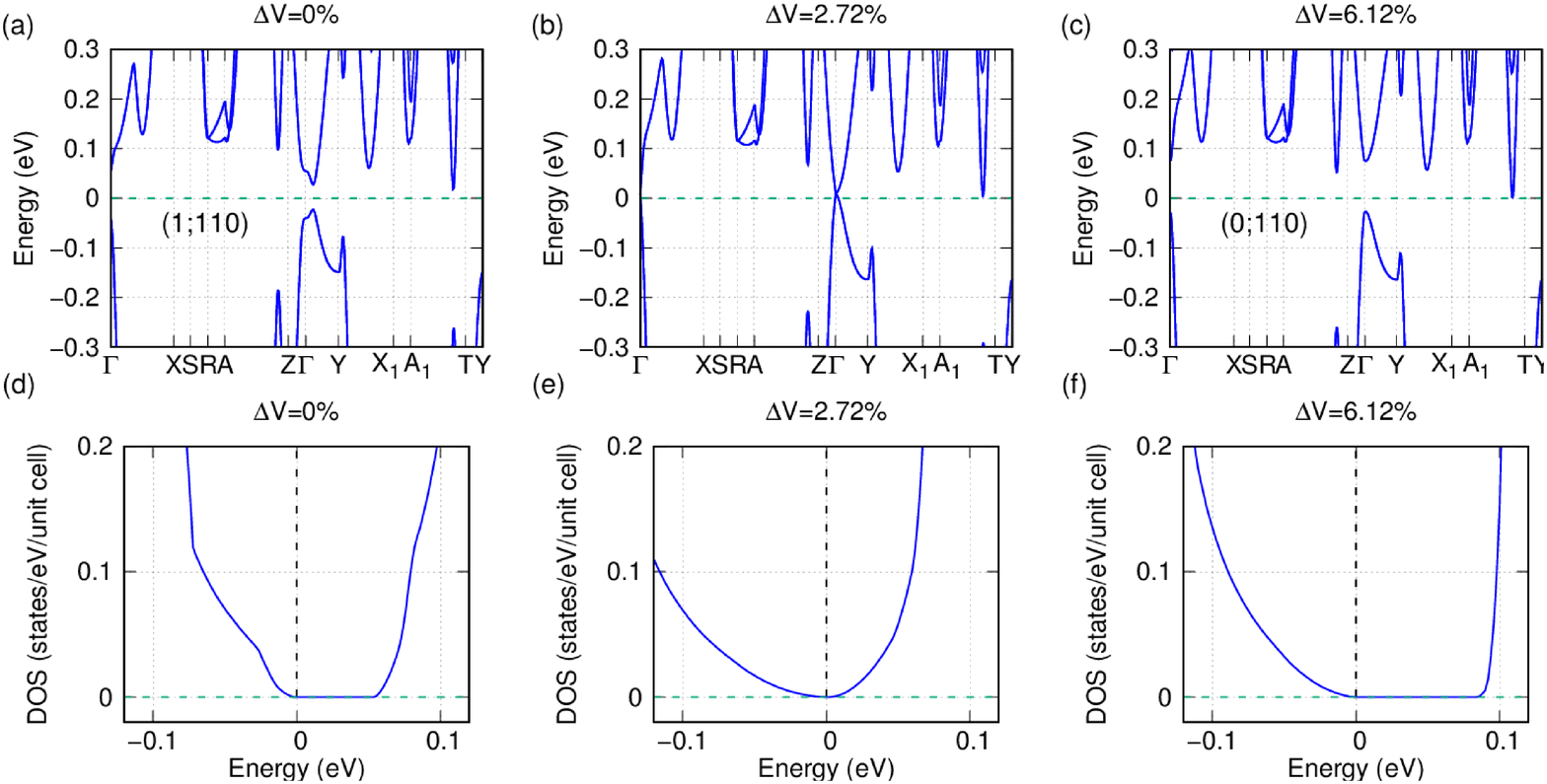}
\caption{\label{fig:str} (Color online) Band structures ({\bf a-c})  and their corresponding DOSs  ({\bf d-f})
with SOC of ZrTe$_5$ under different volumes. The high-symmetry points are given in Fig. 1(b).
Fermi energy is set as 0.}
\end{figure}

\begin{figure}[ht]
\centering
\includegraphics[width=0.8\linewidth]{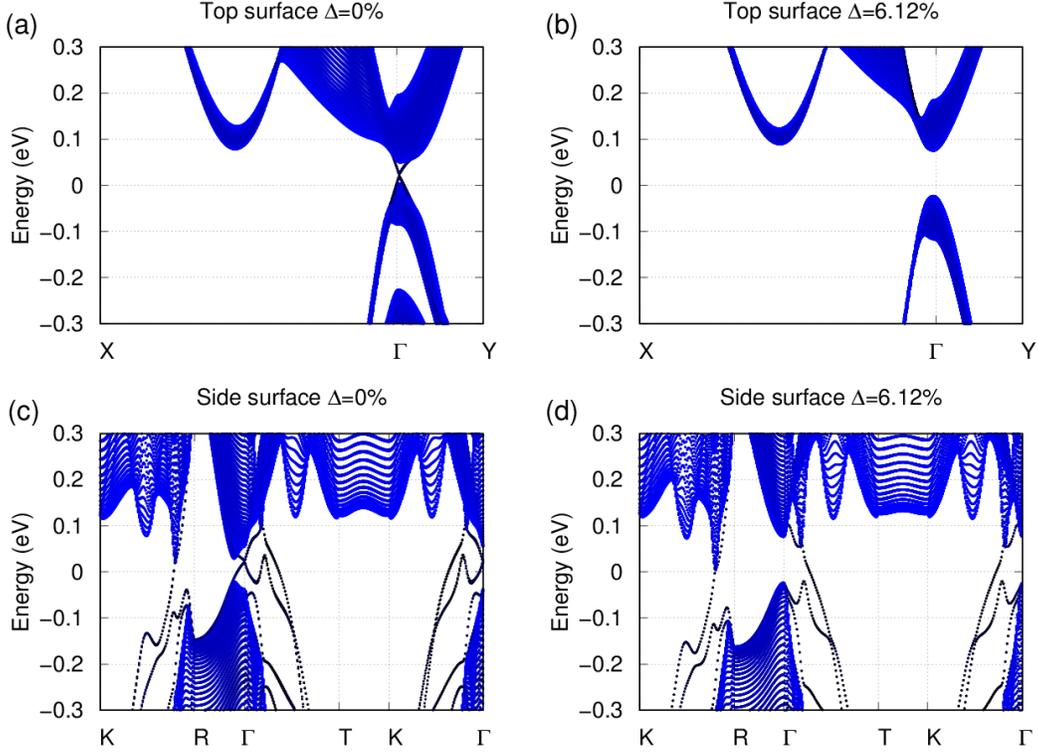}
\caption{\label{fig:str}    (Color online) Calculated surface states of the top surface ($a$-$c$ plane, i.e. cleavage surface)
for the strong (a) and weak (b) TI phase. Calculated surface states of the side surface ($a$-$b$ plane) for the strong
(c) and weak (d) TI phase. Fermi energy is set as 0.  }
\end{figure}

\begin{figure}[ht]
\centering
\includegraphics[width=0.6\linewidth]{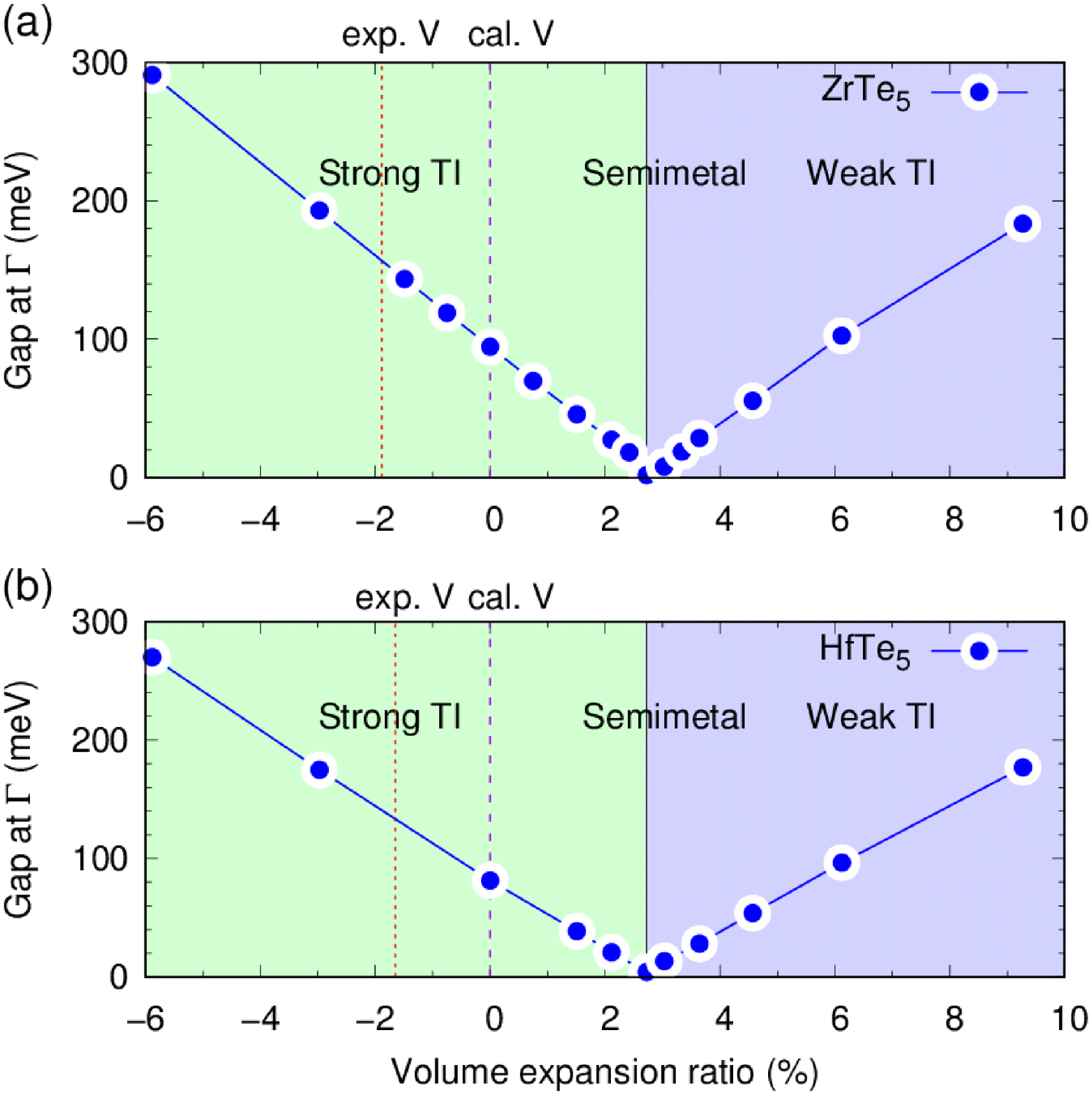}
\caption{\label{fig:str} (Color online)  ({\bf a}) Calculated absolute value of direct band gap at  $\Gamma$ point of ZrTe$_5$
under different volumes.  The light green and  blue  region represent the phases of strong and weak TIs respectively.
The boundary between the strong and weak TI is the semi-metal state.
The red and blue dotted vertical lines represent the experimental volume at 10 K and
calculated one at ground state respectively.
({\bf b}) Same as ({\bf a}) but for HfTe$_5$.
}
\end{figure}

In order to explore the possible topological phase transition in  ZrTe$_5$ ( HfTe$_5$),
we then study their electronic properties under different volumes.  Based on the above optimized structure,
we change volume of the unit cell by hand and then optimize the atom positions and lattice constants
under each variant volumes. This process can simulate the hydrostatic pressure experiments or the thermal
expansion effect due to finite temperatures. It is noted that we did not change the volume drastically and the
system is far away from the region of superconductivity phase under high pressure found in ZrTe$_5$ and HfTe$_5$. ~\cite{sc1,sc2,sc3}
In Fig. 2(a), we present the change of lattice constants $a$, $b$, and $c$ under different volume expansion ratio,
defined as $ (V-V_0)/V_0 \times 100 \%$, where $V_0$ is the unit cell volume at theoretical ground state listed in the Table I.
It is found that all the lattice constants have similar linear dependence on the volume of the unit cell.
But the in-plane lattice constants $a$ and $c$ changes much slower with the volume than that
of the lattice constant $b$,  which indicates the weak inter-layer binding energy along the $b$ direction in ZrTe$_5$.
The parabolic-like relationship between total energy of unit cell and volume is expected and given in Fig. 2 (b). The blue
vertical dotted line represent the experimental volume at low temperature (10 K),
which is nearly 1.9\% smaller than our calculated value.

Then we have calculated the band structures and the DOSs with spin-orbit coupling (SOC) under variant volumes,
and three of them are shown in Fig. 3. The calculated band structure (Fig. 3 (a)) at the ground state volume ($\Delta V=0$)
is similar as previous theoretical computation~\cite{wenghm1}, although we use a different high symmetry k-path.
A clear direct band gap about 94.6 meV at $\Gamma$ point is found in the band structure. Of course
it is also found that the valence band maxima is between the $\Gamma$ and Y point and conduction band minima
is between the A$_1$ and T point. Therefore the indirect band gap is much smaller than the direct one at $\Gamma$ point,
which is about 41.7  meV in Fig. 3(a).
The present of a clear band gap is  confirmed in its corresponding DOS (Fig. 3 (d)).
Our calculated band gap is comparable with the values observed in the previous
experiments, which is 80 or 100 meV. ~\cite{lisc1,wenghm2}
  We also calculated the 3D iso-energy surface (not shown here) of ZrTe$_5$ in the whole
Brillouin zone by the wannier functions  and confirmed again that there is a global band gap in  ZrTe$_5$ when
$\Delta V=0\%$.

 When  ZrTe$_5$ expands from its ground state, the band gap decreases gradually
 until the valence and conduction bands touch each other at a critical volume expansion ratio
 about 2.72\%. Then, a Dirac point  is formed at $\Gamma$ point, which can be
 clearly seen in Fig. 3(b).  This behavior is also confirmed in its corresponding V-shaped
 DOS near the Fermi energy, as shown in Fig. 3(e), which is the feature of Dirac point in band structure.
  It is noted that this Dirac point is 4-fold degenerate since  ZrTe$_5$ has both the space inversion and time reversal symmetry.
 As the crystal continues to expand, the band gap of  ZrTe$_5$ opens again, and ultimately reaches a value
 of about 102.6 (direct) or 27.7 meV (indirect) under a volume expansion 6.12\%. (see Figure 3 (c) and (f)).
  This band gap is also confirmed by the 3D  iso-energy surface of ZrTe$_5$ in the
Brillouin zone.
Therefore from Fig. 3, we can clearly see a transition from a semiconductor to a semi-metal and then
to a semiconductor again in ZrTe$_5$ when it expands.
In order to check whether such a transition is topological or not, we have calculated the $Z_2$
indices under each volume.~\cite{z2pack} It is found that
the  $Z_2$  indices are all (1;110) when the volume expansion is less than 2.72\%, while it is (0;110) when
the volumes expansion is larger than 2.72\%. This definitively confirms that ZrTe$_5$ undergoes a
topological phase transition from a strong TI, to an intermediate Dirac semi-metal state, and finally turns to a weak TI when
its unit cell expands from 0 to 6.12\% in our calculation.
 We noted that our calculated weak indices (110) are different from Weng's calculation~\cite{wenghm1} but same as
Manzoni's ~\cite{manzoni1}  since the weak indices of $Z_2$ depend on the choice of the unit cell. ~\cite{z2pack}

The surface states of ZrTe$_5$ in the strong and weak TI phase have also been calculated based on the wannier functions, shown
in Fig. 4. The surface band structures are very similar as the one presented in Weng's work, ~\cite{wenghm1} since we use
the similar high symmetry k-path in the surface Brillouin zone.  From Fig. 4, we can see that there is a Dirac point at  $\Gamma$
point in top surface's band structure for the ZrTe$_5$ of $\Delta V=0\%$ while it does not for the case of  $\Delta V=6.12\%$.
This key difference confirms again that ZrTe$_5$ is a strong TI when $\Delta V=0\%$  and it becomes a weak TI when $\Delta V=6.12\%$.

The detailed  phase diagrams of such a topological phase transition of  ZrTe$_5$ and  HfTe$_5$ are given
in Fig. 5, in which all the calculated absolute value of direct band gaps at $\Gamma$ point under different volumes are plotted.
In Fig. 5(a) we can find that the band gaps of ZrTe$_5$ decreases linearly as the volume increases from a negative volume expansion
ratio about -6\%, with a rate around -33 meV per 1\% change of volume, where the negative value means a decrease
 of the band gap when the crystal expands.  The band gap disappears at $\Delta V = 2.72\%$.
Then it raises linearly with volume in a similar rate of 28 meV per 1\% change of volume.
Therefore ZrTe$_5$ undergoes a topological transition from a strong TI to a weak TI due to volume expansion.
Such a transition must need a zero-gap intermediate
state, which is the Dirac semi-metal state found at about $\Delta V = 2.72\%$ in our calculation.
Similar phase diagram is also found recently by Manzoni \textit{et al.},~\cite{manzoni1} in which
they present the band gap at $\Gamma$ point as a function of the inter-layer distance, but not the volume of the unit cell.
  It is known that the mono-layer ZrTe$_5$ is a quantum spin Hall insulator.~\cite{wenghm1}
When we stack many mono-layers of ZrTe$_5$ into a 3D bulk ZrTe$_5$ crystal, it would be a 3D strong or weak TIs
which is depending on the strength of coupling between the adjacent layers. From Manzoni's~\cite{manzoni1} and our calculation, it is obvious
that the inter-layer distance is the key factor that causes the transition between the strong and weak TI phases in ZrTe$_5$.
In Fig. 5(b), we also show that HfTe$_5$ has the very similar topological phase transition, with almost
the same transition critical volume expansion ratio at about 2.72\%. The band gap of HfTe$_5$ also
changes linearly as the volume increases with a rate about -31 and 26 meV per 1\% change of
volume in the strong and weak TI region respectively.

\section*{Discussion}

The changing rate of our calculated band gap  is quite significant especially in a small band gap semiconductor material.
Therefore, we can conclude that the electronic properties of ZrTe$_5$  (HfTe$_5$) are indeed very
sensitive to the change of the volume and they are indeed located very close to the boundary between
the strong and weak TI.  Although our optimized and the experimentally measured volume of ZrTe$_5$
 (HfTe$_5$) both indicate that they should be within the strong TI region, we think it still has the possibility
 that in experiment ZrTe$_5$ (HfTe$_5$) can locate in a weak TI region
 due to different growth methods and characterization techniques in experiments.
 According to Fig. 5, it is even possible that  ZrTe$_5$  (HfTe$_5$) can be very close to  the intermediate Dirac semi-metal
 state if it happens that its unit cell has a proper volume expansion ratio, which, however, is quite challenging in experiment.
 Another more possible reason  which can explain the semi-metal behavior found in  experiments is due to the defect and doping,
 which make the ZrTe$_5$ (HfTe$_5$)  being a degenerate semiconductor. In a degenerate semiconductor,
 the Fermi energy is located within the conduction or valence band due to the doping effect, and the crystal
 will behave like a metal. But in this case, the energy gap still exist just below or above the Fermi energy
 and the Dirac point is not needed in the energy gap.
This possibility is verified in a recent experimental work by Shahi \textit{et al.} They found
 that the resistance anomaly of ZrTe$_5$, which was observed in many current experiments,  is due to the Te
 deficiency, while the nearly stoichiometric ZrTe$_5$ single crystal shows the normal semiconducting transport
 behavior.~\cite{bipolar}
 In order to avoid the possible artificial effect induced by the cleavage in both STM and ARPES experiments, we suggest that
nondestructive optical measurements for the existence of a direct band gap at $\Gamma$ point, and its change under
 different temperatures,  in the high quality and stoichiometric single crystals are
 probably useful to elucidate the topological nature in ZrTe$_5$  and HfTe$_5$.

Finally we show the importance of our calculated change rate of band gap. First we
 can roughly estimate the bulk thermal expansion coefficient from experimental lattice constants of
ZrTe$_5$ (Table I) to be about $3.4 \times 10^{-5}  $ K$^{-1}$, which means that the volume will change about 1\%
when the temperature changes from 0 to 300 K, equivalently the band gap of ZrTe$_5$ at $\Gamma$
point will change about -33 meV for strong TI phase or 28 meV for weak TI phase, according to our calculation.
In a recent high-resolution ARPES work,~\cite{yzhang} Zhang \textit{et al.} found a clear and dramatic temperature
dependent band gap in ZrTe$_5$, from which we then can estimate that change rate of observed band gap is
about 26 or 37 meV from 0 to 300 K depending on the methods used in their experiment. These two values are
both well consistent with our calculated result. Moreover, the positive change rate found in the
experiment ~\cite{yzhang}  implies that the ZrTe$_5$ crystal used in their experiment is probably a weak TI
according to our calculated phase diagram in Fig. 4.

In summary, we have studied the band structures of ZrTe$_5$ and HfTe$_5$ at variant volume
by first principles calculations. A clearly volume dependent strong and weak topological phase
transition is found, accompanied by an intermediate Dirac semi-metal state at the boundary between the transition.
The direct band gap of ZrTe$_5$ at $\Gamma$ point changes linearly with the volume, which is -33 meV and 28 meV
in a strong and weak TI phase respectively, if the volume of ZrTe$_5$ increases 1\%,
or equivalently if the temperature increases from 0 to 300 K.
The results for HfTe$_5$ is very similar to those of ZrTe$_5$.
Our calculated results indicate that the electronic properties and topological nature of  ZrTe$_5$ and HfTe$_5$  are
indeed very sensitive to the lattice constants of crystals, which is probably the reason for the divergent experimental results
at present. We suggest that high quality and stoichiometric single crystal with accurate structure refinement at
different temperatures would be helpful to resolve the divergent experimental results in ZrTe$_5$ and HfTe$_5$.

\section*{Methods}

The geometric and electronic properties of ZrTe$_5$ and HfTe$_5$ are calculated by the
 density functional theory in the generalized gradient approximation implemented
 in the Vienna Ab-initio Simulation Package (VASP) code.~\cite{kresse1,kresse2}
 The projected augmented wave method ~\cite{paw2,paw1}  and the van der Waals (vdw) corrected
 optB86b-vdw functional ~\cite{b86b1,b86b2} are used.
 The plane-wave cutoff energy is 300 eV and  the k-point mesh is $8\times 8 \times 4$   in the calculations.
 And a  denser k-point mesh of $24 \times 24\times 12$  is used in the density of state (DOS) calculation.
 Spin-orbit coupling (SOC) is included in the calculation except for the structural optimization.

 The theoretical ground states of ZrTe$_5$ and HfTe$_5$ are obtained by fully optimization
 of the atom positions and lattice constants, until the maximal residual force is less than 0.01 eV/\AA.
 Then we vary and fix the volumes of unit cell and still optimize the atom positions and lattice constants to
study the possible topological transition in ZrTe$_5$ and HfTe$_5$.

The maximally-localized Wannier functions  of ZrTe$_5$  are fitted based on the Zr's $d$ and
Te's $p$ orbitals by the Wannier90 code ~\cite{wannier}  and  then the surface states are calculated by the
WannierTools.~\cite{wt}

\bibliography{reference}

\section*{Acknowledgements }

This work is supported by the National Key R\&D Program of China (2016YFA0201104),
the State Key Program for Basic Research (No. 2015CB659400),  and
the National Science Foundation of China (Nos. 91622122, 11474150 and 11574215).
The use of the computational resources in the High Performance Computing Center of Nanjing University
for this work is also acknowledged.

\section*{Author contributions statement}

Y.B.C., S.H.Y., and J.Z. proposed the idea. Z.J.F., Q.F.L. and J.Z. carried out the calculations,
analysed the results, and plotted the figures. Z.J.F. and J.Z. wrote the mannuscript.
All authors reviewed the manuscript.

\section*{Additional information}

\textbf{Competing financial interests:} The authors declare no competing financial interests.

\end{document}